\documentclass[
  aps,
  pre,
  reprint,
  superscriptaddress,
  longbibliography,
]{revtex4-2}

\usepackage{graphicx}
\usepackage{bm}
\usepackage{hyperref}
\usepackage{physics} 

\begin{document}
\newcommand{\eqname}{Eq.}
\newcommand{\secname}{Sec.}

\title{Energy rectification in active gyroscopic networks under time-periodic modulations}

\author{Zhenghan Liao}
\affiliation{Department of Chemistry, University of Chicago, Chicago, IL, 60637, USA}
\author{Suriyanarayanan Vaikuntanathan}
\email{svaikunt@uchicago.edu}
\affiliation{Department of Chemistry, University of Chicago, Chicago, IL, 60637, USA}
\affiliation{James Franck Institute, University of Chicago, Chicago, IL 60637, USA}

\begin{abstract}
Combinations of gyroscopic forces and nonequilibrium activity has been explored recently in rectifying energy in networks with complex geometries and topologies [Phys. Rev. X 10, 021036]. Based on this previous work, here we study the effect of added time-periodic modulations. Numerical calculations show that the time-modulated network generates net energy transport between sites and the surroundings, even in the absence of any temperature gradients. Combining path integral formulation and diagrammatic expansion, we explain how such anomalous energy transport emerges, and show how the transport pattern in complex networks can be connected to relatively simple local structures.
\end{abstract}

\maketitle

\section{Introduction} \label{sec:intro}

Pioneering studies on energy rectification have shown how energy fluxes can be generated in the absence of temperature biases \cite{Flach2002BrokenSymmetries,Das2002RatchetEnergy,Li2008RatchetingHeat,Ren2010EmergenceControl,Ren2012GeometricHeat,Li2012ColloquiumPhononics,Sabass2017FluctuatingLorentzforcelike,Kanazawa2013HeatConduction,Candido2017MacroscopicViolation,Colombo2019HeatFlux,Zhu2016PersistentDirectional,Zhu2018TheoryManybody,Dubi2011ColloquiumHeat}. 
Such principles can potentially be applied to build nanoscale energetic rectifiers \cite{Li2012ColloquiumPhononics}.
From a theoretical perspective, energy transport is usually associated with phonons, but these collective excitations are more difficult to manipulate compared with single particles \cite{Li2012ColloquiumPhononics,Reimann2002BrownianMotors}. 
Previous studies have exploited opportunities provided by nonlinear interactions \cite{Ren2010EmergenceControl}, athermal baths \cite{Das2002RatchetEnergy}, geometric phases from adiabatic modulations \cite{Ren2012GeometricHeat}, or quantum Floquet systems \cite{Kolodrubetz2018TopologicalFloquetThouless}.
Using a combination of parity-breaking metamaterials and nonequilibrium forcing, our recent work \cite{Liao2020RectificationNonequilibrium} uncovered new rectification principles which manifest as directed energy flows between sites in network systems.
Unlike many previous studies that focused on transport between two terminals which are linked directly \cite{Ren2010EmergenceControl} or through an asymmetric segment \cite{Das2002RatchetEnergy,Li2008RatchetingHeat,Ren2010EmergenceControl}, our setup placed all nodes and their connections on a equal footing \cite{Zhu2016PersistentDirectional,Zhu2018TheoryManybody,Dubi2011ColloquiumHeat}, thus enabled extending rectification studies to networks with complex topologies and geometries. 

Based on our recent work \cite{Liao2020RectificationNonequilibrium}, here we study the effect of added time-periodic modulation. 
Our model system is a class of spring-mass networks where each mass is subject to time-modulated Lorentz force \cite{Nash2015TopologicalMechanics,Mitchell2018AmorphousTopological} and is immersed in an active bath \cite{Hanggi1994ColoredNoise}. Using numerical calculations, we show that the time-modulated system is able to rectify energy fluxes between nodes and the bath.
In other words, our model can act as a many-body energy pump despite the absence of temperature biases.
As a comparison, our previous unmodulated system \cite{Liao2020RectificationNonequilibrium} supports net energy transport between sites but not between sites and baths. The modulation thus expands the toolbox for manipulating energy transport in complex networks.

We capture the numerical results by developing an analytic framework to understand the energy rectification in complex networks under time-periodic modulations.
We first expand the energy transport with respect to the modulation amplitude using the Martin–Siggia–Rose / Janssen–De Dominicis–Peliti (MSR/JDP) path integral formalism \cite{Martin1973StatisticalDynamics,Janssen1976LagrangeanClassical,DeDominicis1978FieldtheoryRenormalization}, which reveals the mechanism for energy pumping through a modulation-induced coupling between different Fourier modes of the response function.
We further perform a diagrammatic expansion with respect to the interaction strength using techniques we developed in Ref.~\cite{Liao2020RectificationNonequilibrium}, which provides a way to understand rectification in complex networks in terms of rectification in local subnetworks. 
Taken together, we demonstrated modulation-induced energy pumping in complex network systems, and developed a theoretical framework to understand the mechanism and organization of the energy rectification.
The rectification principle improves our understanding of energy transport and its control in complex systems.

The remainder of this manuscript is organized as follows.
In \secname~\ref{sec:model}, we introduce our time-modulated active gyroscopic model, provide a microscopic definition for the energy flux, and present numerical results.
In \secname~\ref{sec:theo_overview}-\ref{sec:theo_diagram} we develop a theoretical framework for the energy flux that combines path integral formalism and a diagrammatic approach.
In \secname~\ref{sec:pattern} we utilize the rectification principle to create flux patterns.

\section{Model systems and energy pumping} \label{sec:model}

\begin{figure}[tbp]
	\centering
	\includegraphics[width=0.48\textwidth]{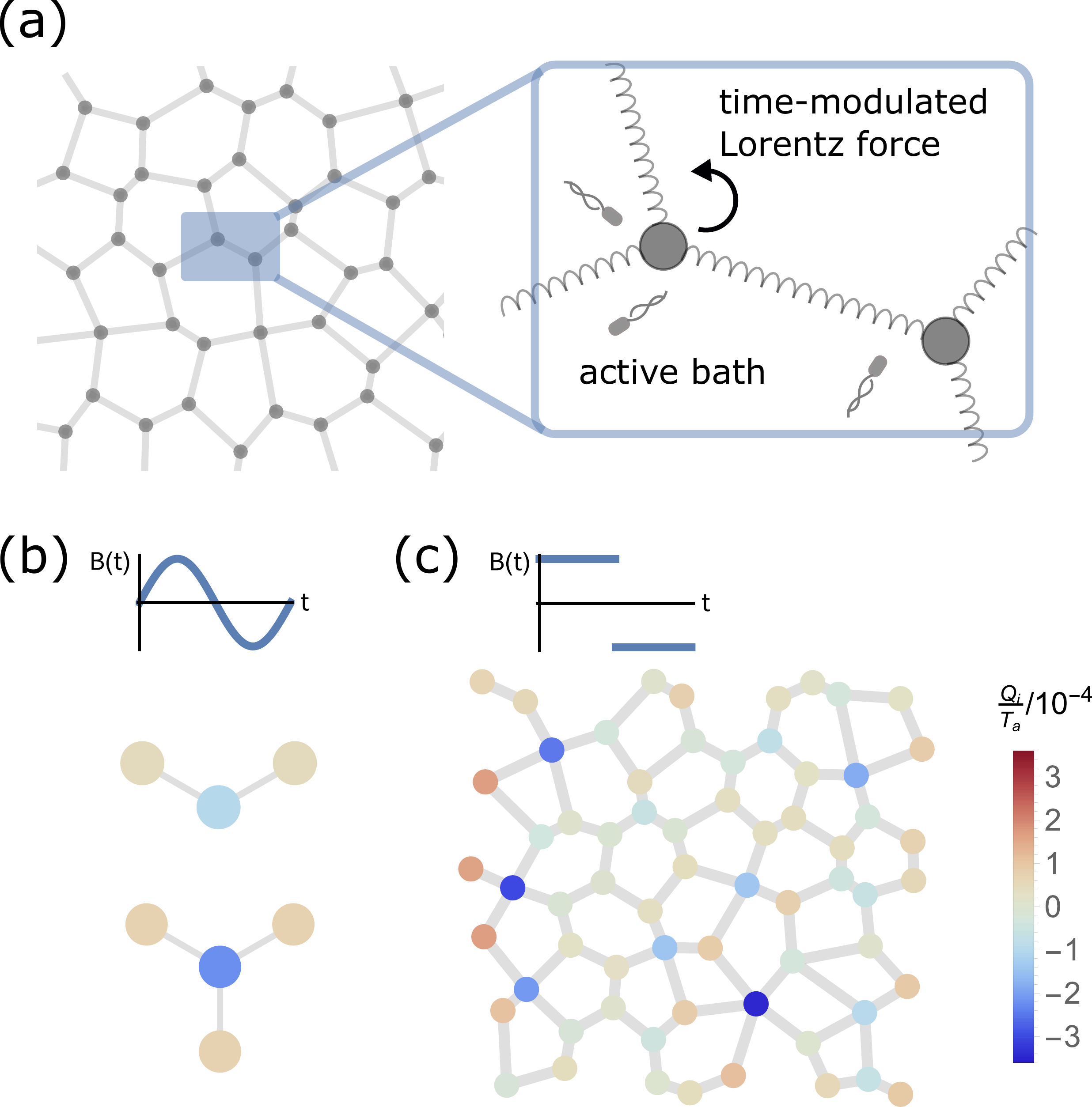}
    \caption{
    The model and energy flux in example networks.
    (a) Schematic of the model, a spring-mass network where each particle is subject to a time-modulated Lorentz-like force and active bath. 
    (b) Energy transferred during each period, $Q_i$, for networks with shape $V$ and $Y$. Positive value corresponds to net energy transferred from the bath to the node. Protocol for $B$-field modulation is $B(t)=\sin 2\pi t/T$, where $T$ is the period of modulation. 
    (c) Energy flux for disordered network subject to a step function protocol $B(t)=1$ if $t<T/2$, $B(t)=-1$ if $t\ge T/2$.
    Numerical calculations were performed with all parameters set to $1$.
    }
    \label{fig:model}
\end{figure}

The equation of motion for our modulated active gyroscopic network model (\figurename~\ref{fig:model}a) reads \cite{Liao2020RectificationNonequilibrium}
\begin{equation} \label{eq:GLE_single}
    m\dot{v}_i = -k_g z_i + \sum_j F_{ji} - \hat{B}(t) A_1 v_i - \gamma v_i + \eta_i.
\end{equation}
We used $z_i \equiv \begin{pmatrix} x_i & y_i \end{pmatrix}^T$ to denote the displacement of particle $i$ from its mechanical equilibrium position. Similarly $v_i$ and $\eta_i$ denote the velocity and the noise. $-k_g z_i$ is an on-site tethering force. The linearized spring force from particle $j$ to $i$ is calculated as $F_{ji} = k (e_{ij}^T z_i + e_{ji}^T z_j) (-e_{ij})$, where $e_{ij}$ is the unit vector that points from the equilibrium position of $i$ to that of $j$. Time-modulated Lorentz force is $-\hat{B}(t) A_1 v_i = -\hat{B}\pmqty{v_{i,y} & -v_{i,x}}^T$, where $\hat{B}=eB$ is the product of the electric charge $e$ and the magnetic field $B$, and the matrix $A_1 \equiv \pmqty{0 & 1 \\ -1 & 0}$.
The last two terms are the friction $-\gamma v_i$ and an Ornstein-Uhlenbeck (OU) colored noise $\eta_i$ \cite{Hanggi1994ColoredNoise} from an active bath. The correlation of the OU colored noise reads
\begin{equation} \label{eq:noise_correlation}
    \expval{\eta_i(t)\eta_{j}^T(t')} = I\delta_{ij}\frac{\gamma T_a}{\tau} e^{-\frac{|t-t'|}{\tau}},
\end{equation}
where $\tau$ is the correlation time, $T_a$ controls the variance of the colored noise, and $I$ is the identity matrix with appropriate dimensions.
The friction $-\gamma v_i$ and the OU noise $\eta_i$ drive the system out of equilibrium via breaking the fluctuation-dissipation relation.
As a result of the periodically modulated $B$-field, the system would reach a time-periodic steady state.

The observable we focus on is the energy transport between particles and baths at the time-periodic steady state. For a system with pairwise interactions and on-site potentials, the energy transferred from bath to particle $i$ in each period $T$, averaged over noise realizations, reads
\begin{equation}
    Q_i = \int_0^T\dd{t} \ev{q_i(t)},
    \quad q_i(t) = -\gamma v_i(t)^Tv_i(t) +v_i(t)^T\eta_i(t). \label{eq:flux_def}
\end{equation}
The first term $(-\gamma v_i^Tv_i)$ measures the energy loss from the particle to the bath due to friction or dissipation. The second term $(v_i^T\eta_i)$ measures the the energy gain for the particle due to fluctuating forcing from the bath.
\eqname~\eqref{eq:flux_def} is derived using stochastic energetics \cite{Sekimoto1997KineticCharacterization,Sekimoto1998LangevinEquation} and a detailed procedure is described in \appendixname~A in Ref.~\cite{Liao2020RectificationNonequilibrium}.

The immediate consequence of time-periodicity is that the total energy transfer during each period is zero, $\sum_{i=1}^N Q_i = 0$, where $N$ is the number of particles in the system.
In nonequilibrium conditions, there seems to be no further constraint on the value of each $Q_i$, thus there is possibility that individual $Q_i$'s are nonzero. Nonzero $Q_i$'s mean that energy is rectified or pumped from some sites to the others.

Starting from the linearized equations \eqname~\eqref{eq:GLE_single}, we numerically solve the time-dependent covariance matrix, from which we calculate $Q_i$ \cite{Gardiner2009ItoCalculus,Ceriotti2010ColoredNoiseThermostats} (\appendixname~\ref{secS:flux_calc}).
\figurename~\ref{fig:model}b-c shows a collection of numerical results for small and larger networks under two example protocols for $B(t)$, a sinusoidal function and a step function. We see that there is energy pumping from some sites to the others. A more detailed description of the average (but not dynamical) picture is as follows, energy is transferred from bath to particles labelled by $Q_i>0$, transmitted through the bonds in the network and released from particles labelled by $Q_i<0$ to their surrounding bath. 
If we were to view this phenomenon from the perspective of conventional temperature-driven transport, we see that although all particles are subject to the same bath or environment, some sites appear \textit{as if} they were hotter ($Q_i>0$) or colder ($Q_i<0$).

Conventional transport theories cannot explain the mechanism of such energy pumping. Moreover, the energy is pumped between multiple sites of the network, contrasting conventional transports between only two terminals. It would be beneficial to develop a theory to capture such kind of transport between multiple sites, and specifically, explain how the pumping depends on the structure of complex networks.

\section{Theory outline: a two-step perturbation strategy} \label{sec:theo_overview}

\begin{figure}[tbp]
	\centering
	\includegraphics[width=0.48\textwidth]{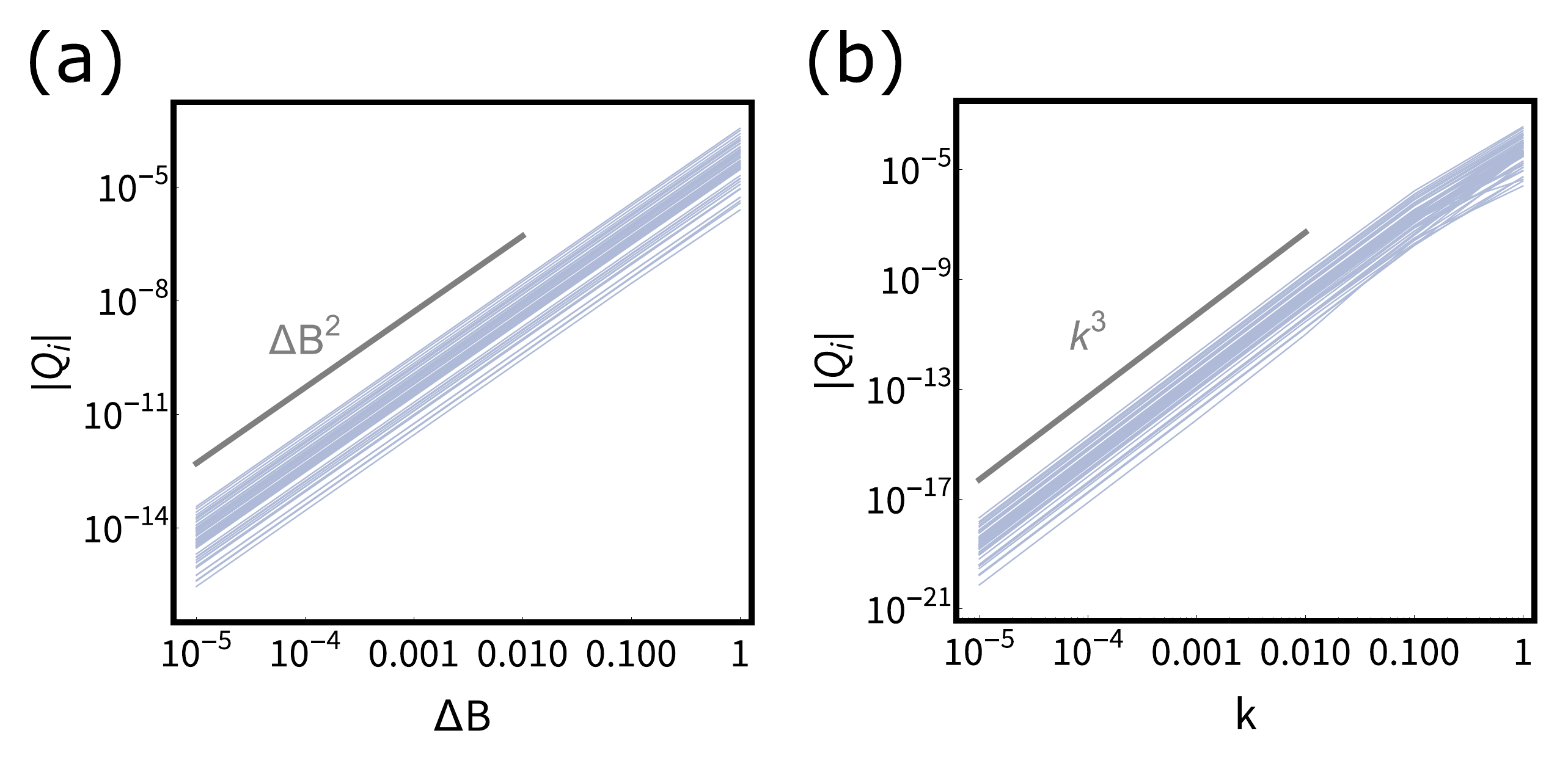}
    \caption{
      Scaling of energy flux with respect to (a) $\Delta B$ and (b) $k$ in for the disordered network in \figurename~\ref{fig:model}c.
      We have separated the modulation $\Delta B(t)$ into an amplitude part $\Delta B$ and a time-dependent part. 
      Each single curve is the scaling for one node.
      Numerical calculations were performed with all other parameters setting to $1$.
    }
    \label{fig:scaling}
\end{figure}

We outline a two-step perturbation theory that aims to explain the emergence of energy fluxes and the connection of fluxes to local properties. Details of each step will be discussed in the next two sections.
This two-step strategy is motivated by our recent work on networks under constant $B$-fields \cite{Liao2020RectificationNonequilibrium}, where we developed a diagrammatic approach as an efficient way to connect transport behavior to local properties. However, the diagrammatic approach cannot directly be applied to the time-modulated case here because its central quantity, the response function, would be invalid. 

To overcome this obstacle, in the first step, we treat the time-modulation as a perturbation and relate the perturbed system to the unperturbed ones. We write the time-modulated $B$-field as $B(t)=B+\Delta B(t)$, where $B$ is a time-independent reference field and $\Delta B(t)$ is a time-periodic modulation with a perturbative amplitude. Using the Martin–Siggia–Rose/Janssen–De Dominicis–Peliti (MSR/JDP) path integral formalism \cite{Martin1973StatisticalDynamics,Janssen1976LagrangeanClassical,DeDominicis1978FieldtheoryRenormalization}, correlators under time-periodic system can be expressed in terms of correlators under a time-invariant reference system. We will see in \eqname~\eqref{eq:dB_2nd} that the response matrix at different Fourier frequencies gets coupled, which is the main mechanism behind the emergence of the energy flux.

In the second step, we are ready to apply a diagrammatic approach similar to Ref.~\cite{Liao2020RectificationNonequilibrium}. We perform an expansion with respect to interactions or the spring constant, and express the energy transfer as intuitive diagrams. These diagrams then enable us to relate the energy flux in complex networks to the structure of local subnetworks.

Numerical results in \figurename~\ref{fig:scaling} show that to the lowest nonvanishing order $Q_i \sim \Delta B(t)^2 k^3$. These observations suggest a goal for analytical efforts, which is to develop an expression to the order of $\Delta B(t)^2 k^3$, explain why lower order terms vanish, and more importantly explore properties of energy pumping on this nonvanishing order. 
We also observed fluxes whose leading order terms are higher than $\Delta B(t)^2$ or $k^3$ in networks with higher symmetries. These networks are special cases and thus are not our focus.

\section{Perturbative expansion in the modulation: a MSR/JDP approach} \label{sec:theo_msr}

\subsection{The MSR/JDP path integral formalism}

The Martin–Siggia–Rose/Janssen–De Dominicis–Peliti (MSR/JDP) path integral formalism \cite{Martin1973StatisticalDynamics,Janssen1976LagrangeanClassical,DeDominicis1978FieldtheoryRenormalization,Hertz2016PathIntegral} is a powerful framework for studying statistical properties, e.g. the average of an observable $O$, of a stochastic trajectory when compared with another trajectory. Applying to our system, the former trajectory is one under modulated $B$-field, in which the average is denoted as $\expval{O}_{B_t}$, and the latter is one under constant $B$-field, in which the average is denoted as $\expval{O}$.
The average $\expval{O}_{B_t}$ can be expressed as a path integral
\begin{equation}
  \begin{split}
    \ev{O}_{B_t} 
    &= \int\mathcal{D}z\mathcal{D}v\mathcal{D}\eta\ O \mathcal{N} \prod_t \delta(\dot{z}-v) \delta(m\dot{v} + \\ &\quad Kz + k_g z + \gamma v + \hat{B}(t)Av - \eta) \mathcal{P}[\eta], \label{eq:msr_obs_full}
  \end{split}
\end{equation}
where the Dirac-$\delta$ functions ensure that the equations of motion are satisfied, $\mathcal{P}[\eta]$ is the probability of the noise, and  $\mathcal{N}$ is the normalization constant.
We have expressed the $N$-particle system using $2N$-dimensional column vectors, e.g. $z=\sum_i \ket{i}\otimes z_i$, where $\ket{i}$ denotes the $2D$ subspace corresponding to particle $i$. The matrix $K$ calculates inter-particle spring forces $F_s$ due to particles' displacements, $F_s=-Kz$. The matrix $A = \sum_i \ket{i}\bra{i}\otimes A_1$.

The modulated $B$-field can be decomposed into a constant part and a (potentially perturbative) time-varying part, $\hat{B}(t) = \hat{B} + \Delta\hat{B}(t)$. Contributions to the path integral from these two parts can be separated via the introduction of an auxiliary field $i u = \sum_i \ket{i}\otimes iu_i$ through $\delta(\hat{B}Av+\Delta\hat{B}(t)Av+\dots) \propto \int \dd{u} e^{-iu^T (\hat{B}Av+\dots)}e^{-iu^T (\Delta\hat{B}(t)Av)}$. 
Further notice that when $\Delta\hat{B}=0$ the path integral \eqname~\eqref{eq:msr_obs_full} reduces to $\ev{O}$. We get
\begin{align}
  \ev{O}_{B_t} = \ev{O e^{-\int\dd{t} \Delta \hat{B}_t iu^T Av}}, \label{eq:msr_obs_simp}
\end{align}
where we have written the time variable in subscripts for simplicity. 
In regimes where the amplitude of $\Delta B(t)$ is small, the right hand side can be expanded,
\begin{equation}
  \begin{split}
    &e^{-\int\dd{s} \Delta \hat{B}_s iu^T Av} = 1 - \int\dd{s} \Delta \hat{B}_s iu_s^T Av_s + \\ &\quad \frac{1}{2}\int\dd{s}\dd{s'} (\Delta \hat{B}_s iu_s^T Av_s)(\Delta \hat{B}_{s'} iu_{s'}^T Av_{s'}) - \dots \label{eq:msr_expand}
  \end{split}    
\end{equation}
In \eqname~\eqref{eq:msr_obs_simp},\eqref{eq:msr_expand}, we have used the MSR/JDP framework to express the average under modulated conditions in terms of some other average under unmodulated conditions.

The observable we are interested in is the energy flux from the bath to the particle, \eqname~\eqref{eq:flux_def}. To account for the site index more conveniently, we use a projection operator $P_i$,
\begin{align}
  P_i = \ketbra{i}{i} \otimes \pmqty{1 & 0 \\ 0 & 1},
\end{align}
and rewrite the flux quantity $q_i(t)$ as
\begin{align}
  q_i(t) = -\gamma (P_iv_t)^T P_iv_t + (P_iv_t)^T P_i\eta_t. \label{eq:flux_proj}
\end{align}

Combining \eqname~\eqref{eq:flux_proj} with the expansion \eqname~\eqref{eq:msr_expand}, we obtain expressions for the pumped energy to different orders in $\Delta B$,
\begin{align}
  \ev{Q_i^{(0)}}_{B_t} &= \int_0^T\dd{t}\ev{q_i(t)}, \label{eq:dB_0th_raw} \\
  \ev{Q_i^{(1)}}_{B_t} &= -\int_0^T\dd{t}\dd{s} \ev{q_i(t) \Delta \hat{B}_s iu_s^T Av_s}, \label{eq:dB_1st_raw} \\
  \ev{Q_i^{(2)}}_{B_t} &= \frac{1}{2}\int_0^T\dd{t}\dd{s}\dd{s'} \Big\langle q_i(t) (\Delta \hat{B}_s iu_s^T Av_s) \nonumber \\ &\qquad (\Delta \hat{B}_{s'} iu_{s'}^T Av_{s'}) \Big\rangle. \label{eq:dB_2nd_raw}
\end{align}
Here we write down the expansion to quadratic order because, as we will show in the next subsection, the zeroth order and the linear order terms vanish.

To calculate the explicit expressions from \eqname~\eqref{eq:dB_0th_raw}-\eqref{eq:dB_2nd_raw}, one requires the evaluation of two-point correlators and multi-point correlators.
The two-point correlators can be expressed in terms of spectral response function $G^+(\omega)$ for the reference system under constant $B$,
\begin{align}
  G^\pm(\omega) = [K + k_gI \pm i\omega(\gamma I + \hat{B}A) - m\omega^2 I]^{-1}. \label{eq:response_G}
\end{align}
The Fourier transform is defined as $\tilde{f}(\omega) = \int_{-\infty}^\infty \dd{t} f(t) e^{-i\omega t}$. The response function describes how the system responds to fluctuations $\tilde{z}(\omega) = G^+(\omega)\tilde{\eta}(\omega)$.
Explicit expressions of relevant two-point correlators are presented in \appendixname~\ref{secS:dB_correlators}.
The multi-point correlators can be written as combinations of two-point correlators via Wick's theorem \cite{Wick1950EvaluationCollision}.

To make our theory more general, we introduce a function $h(\omega)$ to describe a generic noise spectrum
\begin{align}
  \ev{\tilde{\eta}(\omega)\tilde{\eta}(\omega')^T} = 2\gamma T_a h(\omega) 2\pi \delta(\omega+\omega') I.
\end{align}
For white noise, $h(\omega)$ is constant. For the OU colored noise described in \eqname~\eqref{eq:noise_correlation}, $h(\omega) = 1/(1+\omega^2\tau^2)$.

The time-periodic modulation $\Delta \hat{B}(t)$ will be represented by its Fourier series with coefficients $\Delta\tilde{B}_n$,
\begin{align}
  \Delta \hat{B}(t) = \sum_{n=-\infty}^{\infty} \Delta\tilde{B}_n e^{i\omega_n t},\quad \omega_n = \frac{2\pi n}{T},
\end{align}
where $n=\dots, -1, 0, 1, \dots$ and $\Delta\tilde{B}_n = \Delta\tilde{B}^*_{-n}$.

\subsection{The zeroth and the linear order flux vanish} \label{sec:theo_msr_1st}

The zeroth order modulation corresponds to a constant $B$-field. This case has been explored previously, which showed that there is no net energy flux between the bath and the particle,  $\ev{Q_i^{(0)}}_{B_t} = \int_0^T\dd{t}\ev{q_i(t)} = 0$ (\appendixname~C in Ref.~\cite{Liao2020RectificationNonequilibrium}).

One may expect that the linear order flux also vanish, because $\sin$-wave modulation and its opposite, $-\sin$, should result in the same periodic steady state. However, this argument does not account for modulations that consist of multiple sinusoidal waves.
Through explicitly calculating the linear order energy flux, we show that different modes of modulation are decoupled, thus the linear order term also vanish (\appendixname~\ref{secS:dB_1st}).

\subsection{The quadratic order flux explains pumping mechanism} \label{sec:theo_msr_2nd}

There is no \textit{a priori} reason for quadratic order energy flux to vanish.
Starting from the expression for $\ev{Q_i^{(2)}}_{B_t}$, \eqname~\eqref{eq:dB_2nd_raw}, we calculate the six-point correlators and get (details in \appendixname~\ref{secS:dB_2nd})
\begin{equation}
  \begin{split}
  \ev{Q_i^{(2)}}_{B_t}
  &= 4\gamma T_a T \sum_{n=1}^{\infty} |\Delta\tilde{B}_n|^2 \int\frac{\dd{\omega}}{2\pi} \Big\{ \\ &\quad \omega^2(\omega+\omega_n) (h(\omega+\omega_n) - h(\omega)) \\ & \Re[i\tr P_i G^+(\omega) A G^+(\omega+\omega_n) A G^+(-\omega)^T] \Big\}. \label{eq:dB_2nd}
  \end{split}
\end{equation}
This theoretical expression can explain how energy pumping is generated in the presence of the colored noise and the modulation.

The role of the colored noise takes effect through the factor $h(\omega+\omega_n) - h(\omega)$. If the noise spectrum $h(\omega)$ is flat, which corresponds to a white noise, this factor vanishes. Only colored noise with non-flat spectrums can generate a nonzero $\ev{Q_i^{(2)}}_{B_t}$.

The role of modulated $B$-field is to induce couplings between different modes of the response function, which is manifest through the factor $G^+(\omega) A G^+(\omega+\omega_n) A G^+(-\omega)^T$. 
This is in contrast with the unmodulated case where $G^+(\omega)$ at different frequencies are uncoupled, which leads to no pumping \cite{Liao2020RectificationNonequilibrium}. Thus the coupling between different modes is one necessary mechanism for energy pumping in our model.

Another flux property related to the modulation is that contributions from different terms in the Fourier series of $\Delta B(t)$ are independent, which can be seen from the summation $\sum_{n=1}^{\infty} |\Delta\tilde{B}_n|^2 (\cdots)$. As a consequence, we only need to discuss the flux from each mode of $\Delta B(t)$. Then the flux for arbitrary modulation protocols can be obtained by weighted combinations of the individual modes.

\section{Further expansion in the interaction: a diagrammatic approach} \label{sec:theo_diagram}

\subsection{Diagrammatic expansion and two useful properties}

\begin{figure}[tbp]
	\centering
	\includegraphics[width=0.48\textwidth]{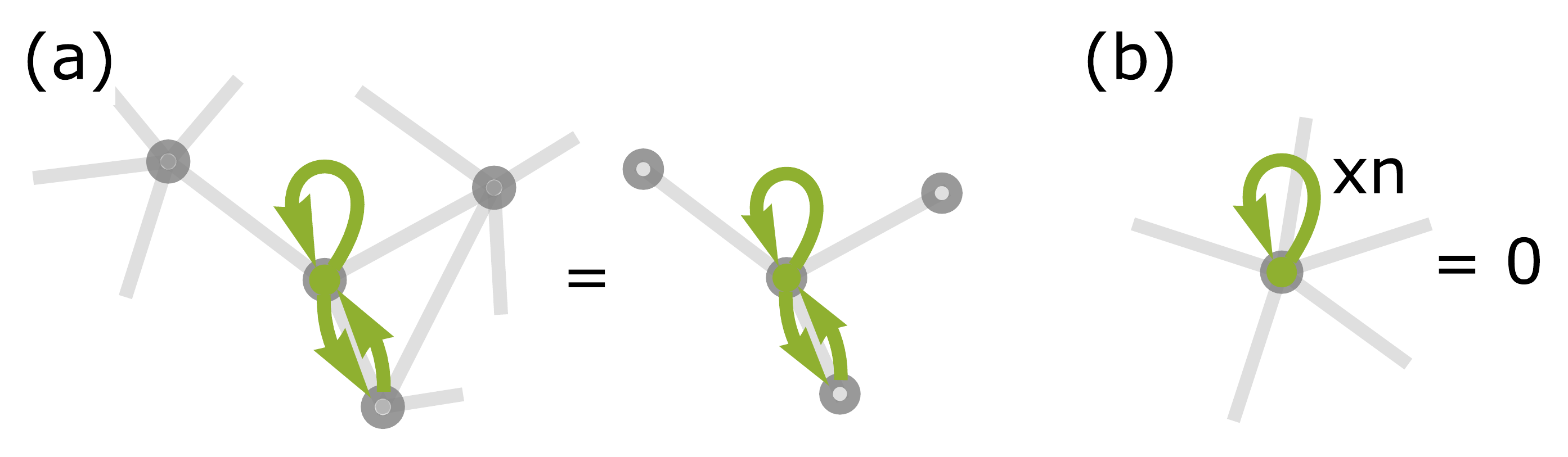}
    \caption{
      Two useful properties of the diagrams. The starting and ending site is labeled in green.
      (a) If there is no loop on a site, then the diagram is equivalent to a simplified one where all other branches on the site are removed.
      (b) Diagrams consisting of only loops vanish. The symbol $n$ in the figure represents an arbitrary number.
    }
    \label{fig:diagram_properties}
\end{figure}

Having explained the mechanism of energy pumping, we now study detailed properties of the flux, in particular, the relationship between flux pattern in a complex network and the structure of local subnetworks.
When interactions are weak, it can be expected that the flux for a node mainly depends on its immediate surroundings. The diagrammtic approach provides a tool to explicitly find such dependence.

Starting from the expression \eqname~\eqref{eq:dB_2nd}, we expand the response functions with respect to small $k$ into products of the noninteracting part (matrix $\eval{G^+}_{k=0}$) and the interacting part (matrix $K$). Due to the pairwise spring-mediated interactions, the matrix $K$ has a block structure, which depends on the topology and the geometry of the network. Further expansion based on the blocks results in terms that can be pictorially represented as diagrams and are closely related to the network structure.
Diagrams for the energy flux between site $i$ and the bath are paths that start from $i$, iteratively step to bonded neighbors or to the site itself, finally ends at site $i$. Diagrams with $|l|$ steps correspond to mathematical expression that are on the order of $k^{|l|}$, which we will denote as $|l|$'s order diagrams. 
In the small-$k$ regime a lower-order diagram contributes more to the flux.
The mathematical expression corresponding to each diagram is lengthy, which we present in \appendixname~\ref{secS:diagram_method}. 

We point out two useful properties of the diagrams. The first property is that if there is no loop on a site, then the diagram is equivalent to a simplified one where all other branches on the site are removed (\figurename~\ref{fig:diagram_properties}a, \appendixname~\ref{secS:diagram_method}). The second property is that for diagrams consisting of solely loops, their value vanish (\figurename~\ref{fig:diagram_properties}b, \appendixname~\ref{secS:diagram_loops}).
As we saw from numerical results in \figurename~\ref{fig:scaling}, energy fluxes scale as $k^3$. Using the two properties described above, we will show in the following two subsections why lower order diagrams vanish and how the third order diagrams reveal an explicit relationship between fluxes in complex networks and local structures.

\subsection{The first and the second order diagrams vanish}

\begin{figure}[tbp]
	\centering
	\includegraphics[width=0.48\textwidth]{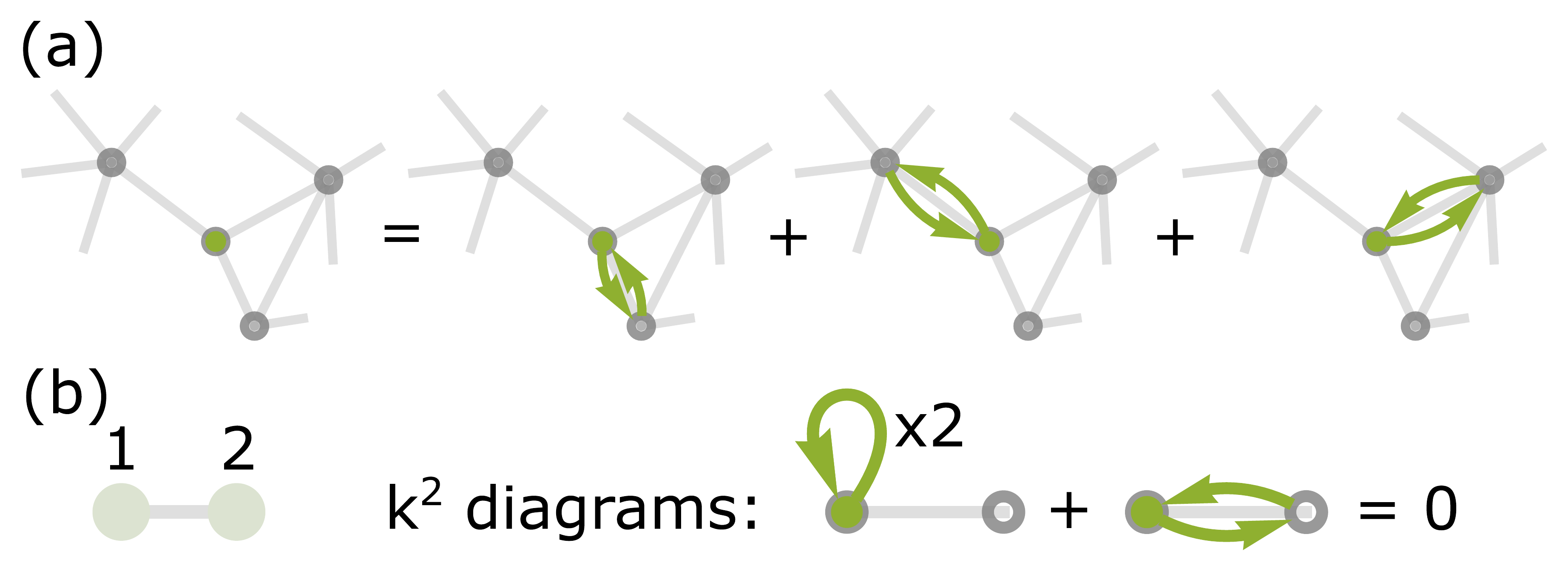}
    \caption{
      Second order diagrams vanish.
      (a) Diagrammatic expansion of the energy flux on the order of $k^2$. Diagrams with only loops are eliminated. From property one, these $i\to j\to i$ diagrams are equivalent to the $1\to 2\to 1$ diagram in a two-node network.
      (b) In the two-node network, the sum of diagram $1\to 1\to 1$ and diagram $1\to 2\to 1$ is zero. Using property two, the diagram $1\to 2\to 1$ vanish.
    }
    \label{fig:diagram_order_k2}
\end{figure}

The first order diagrams means those with only one step. The only possible first order diagrams are those with one loop on the node. According to the second property, all first order diagrams vanish.

Second order diagrams also vanish for the following reason. The second order diagrams for site $i$ have the form $i\to j\to i$, where $j$'s are the bonded neighbors (\figurename~\ref{fig:diagram_order_k2}a, diagrams with only loops are eliminated). Using property one, these diagrams are equivalent to the diagram $1\to 2\to 1$ in a network that consists of only two nodes, $1$ and $2$. From symmetry, the flux in the two-node network is zero. At the $k^2$ level, this means that the sum of a looped diagram and a $1\to 2\to 1$ diagram is zero (\figurename~\ref{fig:diagram_order_k2}b). Thus, the $1\to 2\to 1$ diagram and equivalently $i\to j\to i$ diagrams vanish.

\subsection{The third order diagrams reveal connection between flux and local properties} \label{sec:diagram_k3}

\begin{figure*}[tbp]
	\centering
	\includegraphics[width=0.9\textwidth]{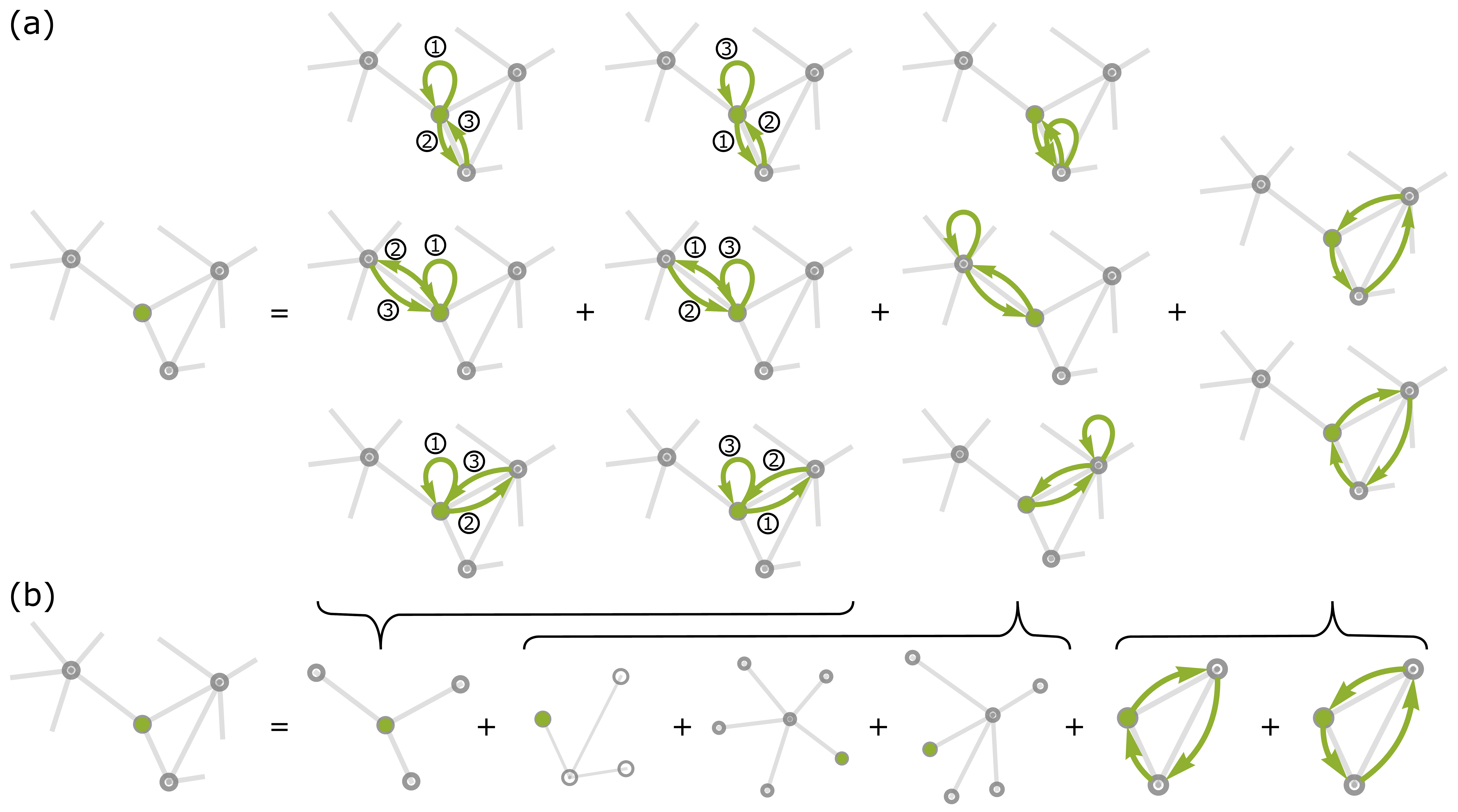}
    \caption{
      Third order diagrams for a node in arbitrary networks. 
      (a) All third order diagrams for the node labeled in green. The number in circles labels the order in the path.
      (b) These diagrams or their partial sums can be classified into three groups and then simplified. The third order flux of the labeled node is equal to the sum of fluxes of corresponding nodes in trimmed subnetworks and fluxes of triangle diagrams.
    }
    \label{fig:diagram_order_k3}
\end{figure*}

The third order diagrams do not vanish in general. Investigation of the third order diagrams shows how the flux in a complex network can be represented using local properties.

In \figurename~\ref{fig:diagram_order_k3}a, we write down all diagrams for a node in arbitrary networks.
The network fragment in \figurename~\ref{fig:diagram_order_k3}a is representative of all possible connections surrounding a node $i$, which include bondings between $i$ and its neighbors $j$'s, bondings between two of its neighbors, and bondings between its neighbors and other nodes in the network.
The flux on the generic node $i$ from the third order equals to the sum of all third order diagrams. 

All third order diagrams can be classified into three classes and then simplified using property one.
The first class of diagrams contain a loop on node $i$, such as $i\to i\to j\to i$, $i\to j\to i\to i$. The sum of all class one diagrams is equal to the third order flux of node $i$ in a trimmed subnetwork centered around $i$, where all neighbors of $j$ and all connections between $j$'s are removed.
The second class of diagrams contain a loop on node $j$, e.g. $i\to j\to j\to i$. Each diagram in class two is equal to the flux of node $i$ in a trimmed subnetwork centered around $j$.
The third class of diagrams are triangles that contain arrows between bonded $j$'s, e.g. $i\to j_1\to j_2\to i$.
The above classification shows an explicit relation between fluxed in complex networks and local properties, which is depicted in \figurename~\ref{fig:diagram_order_k3}b.

\begin{figure}[tbp]
	\centering
	\includegraphics[width=0.48\textwidth]{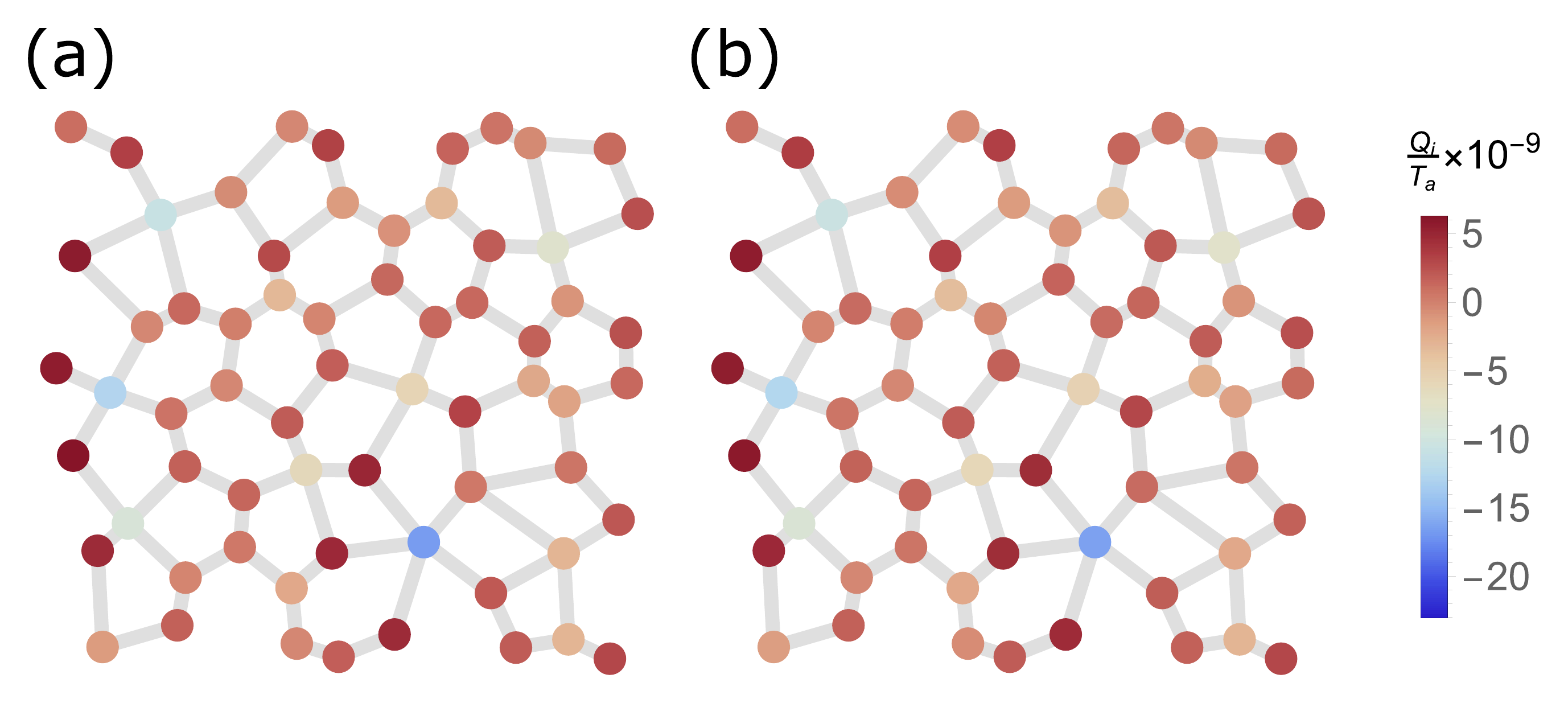}
    \caption{
      Energy fluxes reconstructed from local subnetworks match well with the original ones.
      (a) Fluxes calculated from the full network.
      (b) Fluxes calculated from local subnetworks then combined according to reconstruction rules.
      Parameters are set to $1$ except that $\Delta \hat{B}=0.1, k=0.1$.
    }
    \label{fig:local_reconstruction}
\end{figure}

If a network does not contain any triangular connections, then its flux can be obtained by summing over trimmed subnetworks of the first two classes. A consequence is that we can simply reconstruct the flux in large-scale networks from small subnetworks. \figurename~\ref{fig:local_reconstruction} is a numerical demonstration that such reconstructed fluxes match well with the original ones.

\section{Utilizing local building blocks to create complex patterns of energy transport} \label{sec:pattern}

\begin{figure}[tbp]
	\centering
	\includegraphics[width=0.48\textwidth]{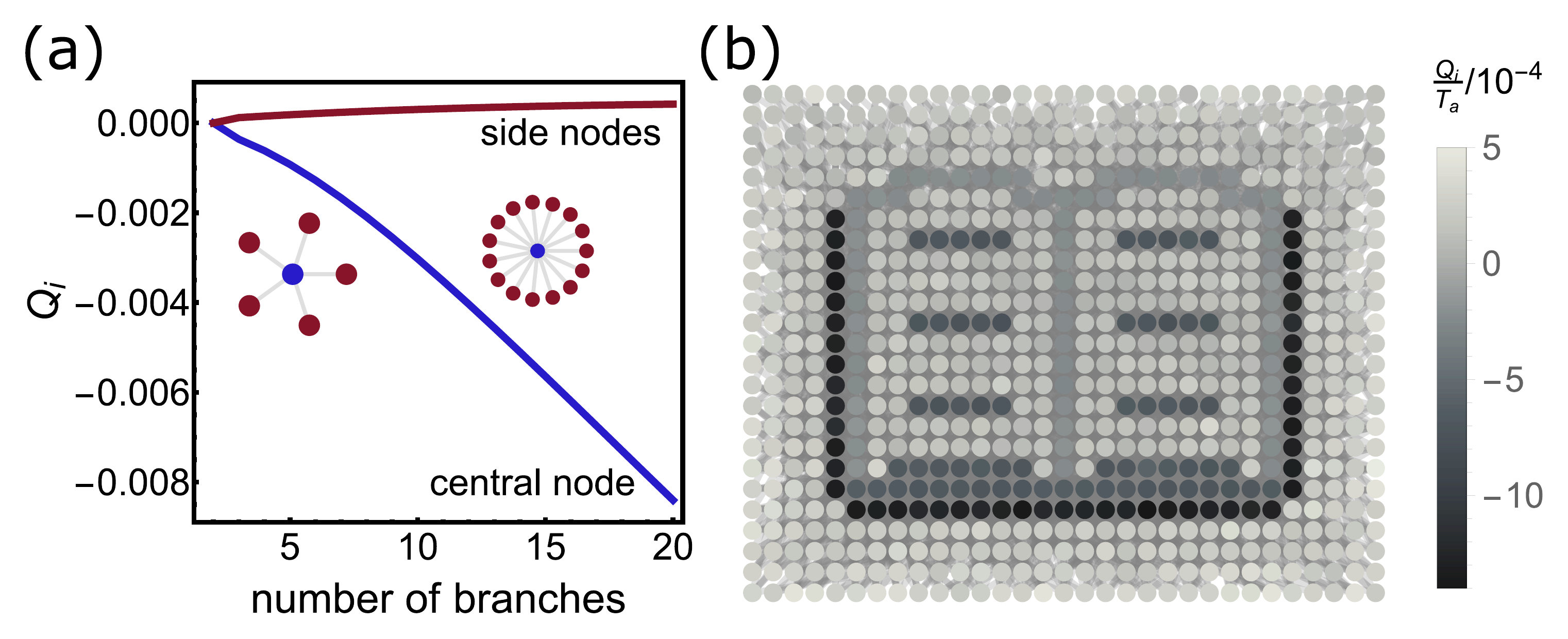}
    \caption{
      Creating target flux patterns by exploiting the connection between flux in a network and in its local subnetworks.
      (a) Flux of network with branches.
      (b) Constructed network and its flux pattern that mimics a grayscaled book.
      The modulation is a step-function modulation where the period is taken to infinite.
      All fluxes are calculated from direct numerical methods. All parameters are set to $1$ except that $k=0.05$.
    }
    \label{fig:pattern}
\end{figure}

The connection between flux in a network and in its local subnetworks can be exploited to create complex patterns of energy transport. 
If we assume that the energy in the bath diffuses slowly, the energy fluxes then can lead to temperature change, which means that our setup could potentially be used to engineer temperature inhomogeneities using homogeneous modulations.

The objective can be posed as follows, given a grid of unconnected nodes and a target pattern, design connections between the nodes such that the consequent flux pattern matches a target one.
From the relation between flux in complex networks and its local structures (\secname~\ref{sec:diagram_k3}), we can inversely use the local subnetworks as building blocks.
The building blocks are networks with one central node and a number of evenly-separated branches. \figurename~\ref{fig:pattern}a shows that the flux of the central node increases in its magnitude as the number of branches increases.
Based on the observed relation between the flux and the number of branches, we can create connections simply by considering the difference between the degree of a node and degrees of its neighbors.
Consider a target line drawing that consists of white background and darker lines, we first highlight nodes corresponding to darker pixels in the pattern. Then we create connections between the highlighted nodes and the non-highlighted ones. The degree of a highlighted node is set by the darkness of its corresponding pixel. Connections to non-highlighted nodes are random, but their average degree is smaller than that of the highlighted ones. We avoid connections between highlighted nodes or between unhighlighted nodes, in order to avoid influence from triangle diagrams.
Note that the connections built from this strategy can be long-ranged in space.
In \figurename~\ref{fig:pattern}b we demonstrate the pattern of a grayscaled book constructed through the above strategy. 
It may be possible to achieve a broader range of patterns and/or avoid long-ranged connections using other strategies.

\section{Conclusion} \label{sec:conclusion}

In conclusion, we have constructed an active gyroscopic network model where the $B$-field is modulated in a time-periodic manner. We numerically demonstrated that our model is able to rectify energy transport between nodes and baths in the absence of any temperature biases. Importantly, by combining the MSR/JDP formalism and our diagrammatic approach, we formulate a connection between flux pattern in complex networks and the flux pattern in local subnetworks. Such connection enables us to understand and control energy pumping in arbitrary complex networks.
The combined MSR/JDP and diagrammatic approach can in principle be applied to calculate generic correlators for perturbed linear networks with arbitrary geometry and topology.

\begin{acknowledgements}
S. V. acknowledges support from the National Science Foundation under Grant No. DMR-1848306.
\end{acknowledgements}

\appendix

\section{Numerical calculation of energy flux from time-dependent covariance matrix} \label{secS:flux_calc}

We first consider the evolution of our active gyroscopic network under a constant $B$-field. Then numerically approximate the time-varying $B$-field by discretizing it into short segments, each under a constant $B$.

Under constant $B$-field, the evolution of our system as described in the extended space $X=\{z, v, \eta\}$ obeys linear dynamics. Here the Ornstein-Uhlenbeck colored noise is treated by the evolution \cite{Hanggi1994ColoredNoise}
\begin{equation} \label{eqS:noise_eom}
    \tau \dot{\eta}_i = -\eta_i + \sqrt{2\gamma T_a}\xi_i.
\end{equation}
For a general linear stochastic equation with time-independent drift $\mu$ and diffusion factor $\sigma$,
\begin{equation}
  \dd X = \mu \dd t + \sigma \dd W,
\end{equation}
its steady-state covariance matrix $C=\expval{XX^T}$ can be numerically solved from the matrix equation $-(\mu C + C \mu^T) = \sigma\sigma^T$ \cite{Gardiner2009ItoCalculus,Ceriotti2010ColoredNoiseThermostats}. 
The evolution of the covariance $C(t)$ starting from an initial state $C_0$ reads
\begin{equation}
  C(t) = C + e^{\mu t}(C_0 - C) e^{\mu^T t}. \label{eqS:cov_time}
\end{equation}
From $C(t)$ we can extract the energy flux, $q_i(t)=-\gamma v_i^Tv_i + v_i^T\eta_i$, and subsequently calculate its time integral.

Our numerical procedure to calculate $Q_i$ is as follows. Given a protocol $B(t)$, we discretize it into short segments in time. In each segment the $B$-field is constant and is evaluated at the starting time of that segment. Consequently, the covariance matrix in each segment can be calculated using \eqname~\eqref{eqS:cov_time}. The evolution $C(t)$ for protocol $B(t)$ can thus be solved by combining results from all segments. We choose a starting $C_0$ to be the steady state under constant $B(0)$, evolve $C(t)$ for many periods until $C(nT) - C(nT+T)$ is smaller than target numerical precision, which indicates that the time-periodic steady state is achieved. Then evolve $C(t)$ from this steady state and calculate the pumped energy $Q_i$. The source of numerical errors mainly come from discretization. Numerical calculations are performed using Mathematica with custom code.

\section{Perturbative expansion in modulated B-field}

\subsection{Two-point correlators expressed in terms of the response function} \label{secS:dB_correlators}

Correlators relevant to calculating $\ev{Q_i^{(1)}}_{B_t}$ and $\ev{Q_i^{(2)}}_{B_t}$ involve $v_t, \eta_t, iu_t$ (but not $z_t$). In this appendix, we express the relevant two-point correlators in terms of the response function $G^+(\omega)$.

Correlators that do not involve the auxiliary field $iu_t$ can be calculated directly.
We show an example calculation of the correlator $\ev{v_tv_s^T}$,
\begin{align}
  \ev{v_tv_s^T} 
  &= \int\frac{\dd{\omega}}{2\pi} e^{i\omega (t-s)}(i\omega)(-i\omega) 2\gamma T_a h(\omega) G^+(\omega) G^-(\omega)^T \\
  &= \int\frac{\dd{\omega}}{2\pi} e^{i\omega (t-s)} i\omega T_a h(\omega) (G^+(\omega) - G^{-T}(\omega)) ,
\end{align}
where to reach the second line we have used $G^{-T} - G^+ = 2i\omega \gamma G^+G^{-T}$ \cite{Kundu2011LargeDeviations}.

To calculate correlators that involve the auxiliary field, we first need to review the connection between the auxiliary field and the response of the system.
Consider an unmodulated system that is perturbed by an external force $f(t)$,
\begin{align}
  m\dot{v} = -Kz - \gamma v - \hat{B}Av + \eta + f.
\end{align}
The MSR/JDP result can be obtained by simply replacing $-\Delta \hat{B} Av$ in \eqname~\eqref{eq:msr_obs_simp} by $f$, which reads
\begin{align}
  \ev{O}_f = \ev{O e^{\int\dd{t} iu^T f}}.
\end{align}
The correlators can then be related to the response,
\begin{align}
  \ev{O iu_{i,s}} &= \eval{\frac{\delta}{\delta f_{i,s}}\ev{O}}_{f\to 0}, \\
  \ev{O iu_{i,s} iu_{i',s'}} &= \eval{\frac{\delta}{\delta f_{i,s}\delta f_{i',s'}}\ev{O}}_{f\to 0},
\end{align}
where we have expressed the component of the vector $f, iu$ explicitly.
From the above expressions we see that these correlators are connected to responses to an external perturbation, for which reason the auxiliary field $iu$ is also called a response field.
For our linear reference system, such response can be expressed in terms of the response function $G^+(\omega)$.

The two-point correlators needed to calculate $\ev{Q_i^{(1)}}_{B_t}$ and $\ev{Q_i^{(2)}}_{B_t}$ are summarized as follows
\begin{align}
  \ev{v_tv_s^T} &= \frac{1}{2\gamma}(\ev{v_t\eta_s^T} + \ev{v_s\eta_t^T}^T), \label{eqS:2pt_vv} \\
  \ev{v_t\eta_s^T} &= 2\gamma T_a \int\frac{\dd{\omega}}{2\pi} e^{i\omega (t-s)} i\omega h(\omega) G^+(\omega), \label{eqS:2pt_ve} \\
  \ev{\eta_t\eta_s^T} &= 2\gamma T_a \int\frac{\dd{\omega}}{2\pi} e^{i\omega (t-s)} h(\omega), \label{eqS:2pt_ee} \\
  \ev{v_tiu_s^T} &= \int\frac{\dd{\omega}}{2\pi}e^{i\omega (t-s)} i\omega G^+(\omega), \label{eqS:2pt_vu} \\
  \ev{\eta_tiu_s^T} &= 0, \label{eqS:2pt_eu} \\
  \ev{iu_tiu_s^T} &= 0. \label{eqS:2pt_uu}
\end{align}

\subsection{Linear order perturbation in modulation} \label{secS:dB_1st}

In this appendix, we derive the linear order energy flux with respect to modulation of the $B$-field. We show that this contribution vanishes.

From \eqname~\eqref{eq:dB_1st_raw},\eqref{eq:flux_proj}, the linear order energy flux reads
\begin{align}
  \ev{q_i(t)^{(1)}}_{B_t} &= -\int\dd{s} \Delta \hat{B}_s \Big\{ \langle -\gamma (P_iv_t)^T P_iv_t iu_s^T Av_s \rangle \nonumber \\ &\qquad + \langle (P_iv_t)^T P_i\eta_t iu_s^T Av_s \rangle \Big\}. \label{eqS:dB_1st_start}
\end{align}

Using the Wick's theorem for four-point correlators,
\begin{align}
  \ev{a^Tbc^Td} &= \tr\ev{ab^T}\tr\ev{cd^T} + \tr\ev{ac^T}\ev{db^T} + \nonumber \\ &\qquad \tr\ev{ad^T}\ev{cb^T}, \label{eqS:wick_4pt}
\end{align}
the first and the second term in \eqname~\eqref{eqS:dB_1st_start} are reduced to,
\begin{align}
  -\gamma\ev{(P_iv_t)^T P_iv_t iu_s^T Av_s} &= -\gamma(\tr P_i\ev{v_tv_t^T}\tr A\ev{v_siu_s^T} \nonumber \\ &\quad + 2\tr P_i\ev{v_tiu_s^T}A\ev{v_sv_t^T}), \\
  \ev{(P_iv_t)^T P_i\eta_t iu_s^T Av_s}
  &= \tr P_i\ev{v_t\eta_t^T}\tr A\ev{v_siu_s^T} \nonumber \\ &\quad + \tr P_i\ev{v_tiu_s^T}A\ev{v_s\eta_t^T}.
\end{align}
The sum of the first terms on both RHS vanishes because
\begin{align}
  (-\gamma\tr P_i\ev{v_tv_t^T}+\tr P_i\ev{v_t\eta_t^T}) = \ev{q_i(t)} = 0.
\end{align}
The sum of the second terms on both RHS can be simplified to $-\tr P_i\ev{v_tiu_s^T}A \ev{v_s\eta_t^T}^T$ using \eqname~\eqref{eqS:2pt_vv}.
Plugging in expressions for correlators presented in \appendixname~\ref{secS:dB_correlators}, we get
\begin{align}
  \ev{q_i(t)^{(1)}}_{B_t} &= 2\gamma T_a \int\frac{\dd{\omega}}{2\pi}\frac{\dd{\omega'}}{2\pi}\dd{s} \Big\{ \Delta \hat{B}_s  e^{i(\omega-\omega')(t-s)} \nonumber \\ &\quad (i\omega)(i\omega') h(\omega') \tr[P_i G^+(\omega) A G^+(\omega')^T] \Big\}.
\end{align}

Integration over $s$ can be written with the Fourier transform of $\Delta B$,
\begin{align}
  \int\dd{s} \Delta \hat{B}_s  e^{i(\omega-\omega')(t-s)} = \Delta\tilde{B}(\omega-\omega') e^{i(\omega-\omega')t}.
\end{align}

We then integrate over $t$. Since $\Delta \hat{B}(t)$ is a periodic function with period $T$, it can be expanded in discrete Fourier modes,
\begin{align}
  \Delta \hat{B}(t) &= \sum_{n=-\infty}^{\infty} \Delta\tilde{B}_n e^{i\omega_n t},\quad \omega_n = \frac{2\pi n}{T}, \\
  \Delta\tilde{B}(\omega) &= \sum_n \Delta\tilde{B}_n 2\pi\delta(\omega-\omega_n), \label{eqS:B_omega_discrete}
\end{align}
with the property $\Delta\tilde{B}_n = \Delta\tilde{B}^*_{-n}$.
The integration over $t$ reads
\begin{align}
  \int_0^T\dd{t} e^{i\omega_n t} &= 
  \begin{cases}
    T, & \text{if } \omega_n = 0 \\
    \frac{1}{i\omega_n}(e^{i\omega_n T}-1) = 0, & \text{if } \omega_n \neq 0
  \end{cases} \\
  &= T\delta_{n,0}. \label{eqS:delta}
\end{align}

We introduce an auxiliary function
\begin{align}
  f_1(\omega, \omega') = 2\gamma T_a (i\omega)(i\omega') h(\omega') \tr[P_i G^+(\omega) A G^+(\omega')^T].
\end{align}
The linear order energy flux at time instant $t$ reads
\begin{align}
  \ev{q_i(t)^{(1)}}_{B_t} &= \int\frac{\dd{\omega}}{2\pi}\frac{\dd{\omega'}}{2\pi}\Delta\tilde{B}(\omega - \omega') e^{i(\omega-\omega')t} f_1(\omega, \omega') \\
  &= \sum_n \int\frac{\dd{\omega}}{2\pi}\Delta\tilde{B}_n e^{i\omega_n t} f_1(\omega, \omega-\omega_n),
\end{align}
which shows that different modulation modes, $\Delta\tilde{B}_n$, are decoupled.

After time integration the result reads
\begin{align}
  \ev{Q_i^{(1)}}_{B_t} &= \int_0^T\dd{t} \ev{q_i(t)^{(1)}}_{B_t} \\
  &= T \int\frac{\dd{\omega}}{2\pi}\Delta\tilde{B}_0 f_1(\omega, \omega) \\
  &= 2\gamma T_aT \int\frac{\dd{\omega}}{2\pi} \Big\{ \Delta\tilde{B}_0 h(\omega) (i\omega)^2 \nonumber \\ 
  &\quad \tr[P_i G^+(\omega) A G^+(\omega)^T] \Big\} = 0. \label{eq:dB_1st}
\end{align}

This result shows that the only contribution is the zero-frequency mode of $\Delta \hat{B}(t)$, thus the flux should vanish.
The mathematical proof is as follows, since $G^+(\omega)^TP_iG^+(\omega)$ is a symmetric matrix and $A$ is an antisymmetric matrix, the trace of their product is zero.

\subsection{Quadratic order perturbation in modulation} \label{secS:dB_2nd}

In this appendix, we derive the expression for the quadratic order energy flux with respect to modulation of the $B$-field, \eqname~\eqref{eq:dB_2nd} in the main text. We also perform sanity checks that the energy balance is satisfied and that flux vanishes if the modulation is constant.

We start from expressions \eqname~\eqref{eq:dB_2nd_raw},\eqref{eq:flux_proj}, and get the quadratic order energy flux at time $t$,
\begin{equation}
  \begin{split}
  \ev{q_i(t)^{(2)}}_{B_t} &=  \frac{1}{2}\int\dd{s}\dd{s'} \Big\langle (-\gamma (P_iv_t)^T P_iv_t + (P_iv_t)^T P_i\eta_t) \\ &\quad (\Delta B_s iu_s^T Av_s)(\Delta B_{s'} iu_{s'}^T Av_{s'}) \Big\rangle.
  \end{split}
\end{equation}
This expression involves six-point correlators, which emit 15 terms using the Wick's theorem.
However, many of these terms will turn out to vanish, which greatly simplifies the calculation. 

Our first task is to identify these vanishing terms.
The quadratic order perturbation can be expanded as
\begin{equation}
  \begin{split}
    &\int\dd{t}\ev{q_i(t)^{(2)}}_{B_t} = \frac{1}{2} \int\dd{t}\dd{s}\dd{s'} \Delta \hat{B}_s \Delta \hat{B}_{s'} \Big [ \ev{q_i(t)} \\ &\quad \ev{iu_s^T Av_s} \ev{iu_{s'}^T Av_{s'}} + \ev{q_i(t) iu_s^T Av_s}_c \ev{iu_{s'}^T Av_{s'}} + \\ &\quad \ev{iu_s^T Av_s} \ev{q_i(t) iu_{s'}^T Av_{s'}}_c + \ev{q_i(t) iu_s^T Av_s iu_{s'}^T Av_{s'}}_c \Big ], \label{eqS:dB_2nd_groups}
  \end{split}
  \end{equation}
where subscript ``c'' means the terms are ``connected" inside the same trace.
The first term vanishes due to $\ev{q_i}=0$. The second and the third term vanish due to $\int\dd{t}\ev{q_i(t)^{(1)}}_{B_t} = 0$.
Now we only need to consider the last term which involves trace connecting all six points.
These terms have the form
\begin{equation} \label{eqS:wick_6pt_connected}
  \begin{split}
  &\ev{a^Tbc^Tde^Tf}_c
  = \tr\ev{ac^T}\ev{df^T}\ev{eb^T} + \\ &\quad \tr\ev{ac^T}\ev{de^T}\ev{fb^T} + \tr\ev{ad^T}\ev{cf^T}\ev{eb^T} + \\ &\quad \tr\ev{ad^T}\ev{ce^T}\ev{fb^T} + \tr\ev{ae^T}\ev{fd^T}\ev{cb^T} + \\ &\quad \tr\ev{ae^T}\ev{fc^T}\ev{db^T} + \tr\ev{af^T}\ev{ed^T}\ev{cb^T} + \\ &\quad \tr\ev{af^T}\ev{ec^T}\ev{db^T}.
  \end{split}
\end{equation}
Applying the above form and notice that some two-point correlators are zero, the expression for the quadratic order energy flux \eqname~\eqref{eqS:dB_2nd_groups} simplifies to
\begin{equation}
  \begin{split}
    &\ev{q_i(t)^{(2)}}_{B_t} 
    = \int\dd{s}\dd{s'} \Delta B_s \Delta B_{s'} \Big\{ \\ &\quad \tr \Big[P_i\ev{v_tiu_s^T}A\ev{v_s\eta_{s'}^T} A\ev{iu_{s'}v_t^T}\Big] -  \\ &\quad \tr \Big[P_i\ev{v_tiu_s^T}A\ev{v_siu_{s'}^T}A\ev{v_t\eta_{s'}^T}^T \Big]\Big\}. \label{eqS:q2_simp}
  \end{split}    
\end{equation}

We next plug in explicit expressions for the two-point correlators \eqname~\eqref{eqS:2pt_vv}-\eqref{eqS:2pt_uu} and integrate over $s,s'$ and $t$. We get
\begin{equation}
  \begin{split}
    \ev{Q_i^{(2)}}_{B_t}
    &= 2\gamma T_a T \sum_{n=-\infty}^{\infty} |\Delta\tilde{B}_n|^2 \int\frac{\dd{\omega}}{2\pi} \Big\{ \\ &\quad i\omega^2(\omega-\omega_n) (h(\omega-\omega_n) - h(\omega)) \\ &\quad \tr P_i G^+(\omega) A G^+(\omega-\omega_n) A G^+(-\omega)^T \Big\}. \label{eqS:Q2_simp}
  \end{split}    
\end{equation}
It can be shown that the $\omega_n$ term and the $-\omega_n$ term form a complex conjugate pair. From this property and \eqname~\eqref{eqS:Q2_simp}, we reach the final expression for the quadratic order energy flux \eqname~\eqref{eq:dB_2nd} in the main text.

\section{Further perturbative expansion in interaction}

\subsection{Procedure and result of the diagrammatic approach} \label{secS:diagram_method}

The diagrammatic approach is built on an expansion of the response function. We first review the diagrammatic expansion of a single response function \cite{Liao2020RectificationNonequilibrium}, then combine the three response functions and other parts in \eqname~\eqref{eq:dB_2nd} or \eqname~\eqref{eqS:Q2_simp} to get the diagrammatic expression for the energy flux.

In the small-$k$ regime, the response function $G^+(\omega)$ (\eqname~\eqref{eq:response_G}) can be expanded as
\begin{equation}
  G^+ = \frac{1}{(G^+|_{k=0})^{-1}+K} = \sum_{|l|=0} G^+|_{k=0} \big [(-K)G^+|_{k=0}\big ]^{|l|} \label{eqS:response_expand_k}
\end{equation}
The noninteracting part $G^+|_{k=0}$ is block diagonal, $G^+|_{k=0} = \sum_i \ket{i}\bra{i}\otimes g^+(\omega)$. Here $g^+(\omega)$ is the $2\times 2$ response matrix for a single noninteracting node, which manifests as a rotation matrix of a complex angle $\alpha_\omega$,
\begin{align}
  g^+(\omega) &= \frac{1}{k_{0,\omega}} (I \cos\alpha_\omega - A_1 \sin\alpha_\omega), \\
  k_{0,\omega} &= \sqrt{(k_g+i\omega \gamma - m\omega^2)^2 - (\omega \hat{B})^2}, \\
  \cos\alpha_\omega &= \frac{1}{k_{0,\omega}} (k_g+i\omega \gamma - m\omega^2), \\
  \sin\alpha_\omega &= \frac{1}{k_{0,\omega}} i\omega \hat{B}.
\end{align}
The interacting part $K$ consists of blocks
\begin{align}
  (-K)_{ii} &= \bra{i}(-K)\ket{i} = \sum_{j, j\neq i} (-e_{ij}e_{ij}^T), \label{eqS:mKs_ii} \\
  (-K)_{ji} &= \bra{j}(-K)\ket{i} = e_{ij}e_{ij}^T, \label{eqS:mKs_ji}
\end{align}
where $e_{ij}$ denotes the unit vector that points from the equilibrium position of $i$ to that of $j$.

We insert resolution of identity $I = \sum_i\ket{i}\bra{i}$ into the expansion \eqname~\eqref{eqS:response_expand_k}. As an example,
\begin{equation}
  \begin{split}
    & \bra{i} G^+|_{k=0} (-K) G^+|_{k=0} (-K) G^+|_{k=0}\ket{j} = \\ &\quad \sum_m g^+(\omega) (-K)_{im} g^+(\omega) (-K)_{mj} g^+(\omega).
  \end{split}  
\end{equation}
For block $(-K)_{im}$ to be nonzero, either site $i$ and site $m$ are bonded, or $m=i$. Likewise for block $(-K)_{mj}$.
These constraints on path $i\to m\to j$ can be conveniently addressed using diagrams: first draw the network, label the nodes $i$ and $j$, then identify nodes $m$'s that satisfy the constraints.

Now we apply the diagrammatic approach to energy flux for site $i$, \eqname~\eqref{eq:dB_2nd} or \eqname~\eqref{eqS:Q2_simp}. Each term in the expansion of the energy flux can be represented as a diagram, or a path $l: i=l_0\to l_1\to \cdots \to l_{|l|-1}\to l_{|l|} = i$, where $|l|$ is the length of the path. Consecutive nodes in the path either has to be bonded or they are the same node. The path has to start and end at node $i$ because the existence of the projection operator $P_i$.
The three $G^+$'s dictates that path $l$ needs to be partitioned into three segments with lengths $\{|l|_1,|l|_2,|l|_3\}$ ($|l|_1+|l|_2+|l|_3=|l|$), and each segment sets how each $G^+$ is expanded.

Taken together, the diagrammatic expression of $\ev{Q_i^{(2)}}_{B_t}$ can be written as a sum over paths,
\begin{align}
  \ev{Q_i^{(2)}}_{B_t} &= T_a \sum_{n=1}^{\infty} T|\Delta\tilde{B}_n|^2 \sum_l k^{|l|} f_{i,n;l}, \label{eqS:path_sum} \\
  f_{i,n;l} &= \sum_{|l|_1+|l|_2+|l|_3=|l|} f_{i,n;l;|l|_1,|l|_2,|l|_3}. \label{eqS:path_one}
\end{align}
$f_{i,n;l}$ denotes the mathematical expression for path $l$. $f_{i,n;l;|l|_1,|l|_2,|l|_3}$ denotes the expression for partition $\{|l|_1,|l|_2,|l|_3\}$, which reads
\begin{equation}
  \begin{split}
    &f_{i,n;l;|l|_1,|l|_2,|l|_3} 
    = 2\Re \int\frac{\dd{\omega}}{2\pi} \omega(\omega+\omega_n) \Big(h(\omega+\omega_n) \\ &\quad - h(\omega)\Big) \tr\Big\{M[(-K)g^+(\omega)]_{l_{|l|_1+|l|_2}\to\cdots i} A \\ &\quad M[g^+(\omega+\omega_n)(-K)]_{l_{|l|_1}\to\cdots l_{|l|_1+|l|_2}}g^+(\omega+\omega_n) A \\ &\quad M[g^+(-\omega)^T(-K)]_{i\to\cdots l_{|l|_1}} (g^+(-\omega)^T-g^+(\omega)) \Big\}. \label{eqS:path_partition}
  \end{split}
\end{equation}
Symbol $M[\cdot]$ denotes the expression for a segment of the path,
\begin{align}
    &M[(-K)g^+(\omega)]_{l_0\to l_1\to \cdots\to l_n} = (-K)_{l_n,l_{n-1}}g^+(\omega)\cdots \nonumber \\ &\qquad (-K)_{l_2,l_1}g^+(\omega)(-K)_{l_1,l_0}g^+(\omega), \\
    &M[g^+(\omega)(-K)]_{l_0\to l_1\to \cdots\to l_n} = g^+(\omega)(-K)_{l_n,l_{n-1}}\cdots \nonumber \\ &\qquad g^+(\omega)(-K)_{l_2,l_1}g^+(\omega)(-K)_{l_1,l_0}.
\end{align} 

From \eqname~\eqref{eqS:path_sum},\eqref{eqS:path_one},\eqref{eqS:path_partition}, we obtain the procedure to write down energy flux for site $i$ on the order of $k^{|l|}$ as follows.
Firstly, draw all possible closed paths with length $|l|$ that starts from node $i$, iteratively navigates to its bonded neighbors or to itself for $|l|$ steps, and ends at node $i$.
Secondly, for each path $l$, find all partitions $\{|l|_1,|l|_2,|l|_3\}$, and calculate $f_{i,n;l;|l|_1,|l|_2,|l|_3}$ according to \eqname~\eqref{eqS:path_partition}.
Finally, sum up all partitions to get $f_{i,n;l}$ (\eqname~\eqref{eqS:path_one}), then sum up all paths to obtain $\ev{Q_i^{(2)}}$ on the $k^{|l|}$ order (\eqname~\eqref{eqS:path_sum}).

Path $l$ and its corresponding mathematical expression $f_{i,n;l}$ can be presented as diagrams. 
An arrow $i\to j$ in the diagram corresponds to $(-K)_{ji}$ mathematically, and as a result, if $i\neq j$, the contribution from this arrow is independent of the other neighbors of $i$ or $j$ (\eqname~\eqref{eqS:mKs_ji}). If $i=j$, however, neighbors of $i$ cannot be removed because they do affect the value of $i\to i$ through $(-K)_{ii}$ (\eqname~\eqref{eqS:mKs_ii}).
As a result, if a diagram contains no loops on some node $j$, the diagram is equal to a trimmed diagram where we remove all neighbors of $j$ except for those appear in the path.
This basic property helps to simplify the diagrams without explicit calculations of $f_{i,n;l}$.

\subsection{Diagrams that consist of only loops vanish} \label{secS:diagram_loops}

For diagrams with only loops, the expression $M[\cdot]$ simplifies to multiplication of the same matrix. Denoting $(-K)_{ii} = M_i$, $f_{i,n;l;|l|_1,|l|_2,|l|_3}$ reads
\begin{align}
  &f_{i,n;l;|l|_1,|l|_2,|l|_3} = f_{M_i,1}(|l|_1,|l|_2,|l|_3) - f_{M_i,2}(|l|_1,|l|_2,|l|_3), \\
  &f_{M_i,1}(|l|_1,|l|_2,|l|_3) = 2\Re \int\frac{\dd{\omega}}{2\pi} \omega(\omega+\omega_n) \Big(h(\omega+\omega_n) \nonumber \\ &\quad - h(\omega)\Big) \tr \Big\{(M_ig^+(\omega))^{|l|_3} A (g^+(\omega+\omega_n)M_i)^{|l|_2} \nonumber \\ &\quad g^+(\omega+\omega_n) A (g^+(-\omega)^TM_i)^{|l|_1} g^+(-\omega)^T \Big\}, \\
  &f_{M_i,2}(|l|_1,|l|_2,|l|_3) = 2\Re \int\frac{\dd{\omega}}{2\pi} \omega(\omega+\omega_n) \Big(h(\omega+\omega_n)  \nonumber \\ &\quad - h(\omega)\Big) \tr \Big\{(M_ig^+(\omega))^{|l|_3} A (g^+(\omega+\omega_n)M_i)^{|l|_2}  \nonumber \\ &\quad g^+(\omega+\omega_n) A (g^+(-\omega)^TM_i)^{|l|_1} g^+(\omega)\Big\}.
\end{align}

From the above definitions, it is straightforward to prove the following three relations,
\begin{align}
  f_{M_i,1}(|l|_1,|l|_2,|l|_3) &= f_{M_i,2}(|l|_1+1,|l|_2,|l|_3-1),\\
  f_{M_i,1}(|l|_1,|l|_2,0) &= -f_{M_i,1}(|l|_2,|l|_1,0), \\
  f_{M_i,2}(0,|l|_2,|l|_3) &= -f_{M_i,2}(0,|l|_3,|l|_2).
\end{align}
With these relations, $f_{i,n;l}$ can be shown to be zero,
\begin{align}
  f_{i,n;l} &= \sum_{|l|_1+|l|_2+|l|_3=l} \Big\{ f_{M_i,1}(|l|_1,|l|_2,|l|_3) - f_{M_i,2}(|l|_1,|l|_2,|l|_3) \Big\} \\
  &= \sum_{|l|_1=0}^{l-1}\sum_{|l|_2=0}^{l-1-|l|_1} f_{M_i,1}(|l|_1,|l|_2,l-|l|_1-|l|_2) + \nonumber \\
  &\quad \sum_{|l|_1=0}^{l}f_{M_i,1}(|l|_1,l-|l|_1,0) - \sum_{|l|_2=0}^{l}f_{M_i,2}(0,|l|_2,l-|l|_2) \nonumber \\
  &\quad \sum_{|l|_1=1}^{l}\sum_{|l|_2=0}^{l-|l|_1} f_{M_i,2}(|l|_1,|l|_2,l-|l|_1-|l|_2) \nonumber \\
  &= 0.
\end{align}
Thus diagrams consist of only loops vanish.

%

\end{document}